\documentstyle[12pt]{article}

\newcommand{\nc}{\newcommand}
\nc{\beq}{\begin{equation}}
\nc{\eeq}{\end{equation}}
\nc{\beqa}{\begin{eqnarray}}
\nc{\eeqa}{\end{eqnarray}}
\def\DS {D\!\!\!\!/}

\input epsf
\newwrite\ffile\global\newcount\figno \global\figno=1

\def\writedef#1{}
\def\figin{\epsfcheck\figin}\def\figins{\epsfcheck\figins}
\def\epsfcheck{\ifx\epsfbox\UnDeFiNeD
\message{(NO epsf.tex, FIGURES WILL BE IGNORED)}
\gdef\figin##1{\vskip2in}\gdef\figins##1{\hskip.5in}
\else\message{(FIGURES WILL BE INCLUDED)}%
\gdef\figin##1{##1}\gdef\figins##1{##1}\fi}
\def\figinsert{}
\def\ifig#1#2#3{\xdef#1{fig.~\the\figno}
\writedef{#1\leftbracket fig.\noexpand~\the\figno}%
\figinsert\figin{\centerline{#3}}\medskip\centerline{\vbox{\baselineskip12pt
\advance\hsize by -1truein\center\footnotesize{  Fig.~\the\figno.} #2}}
\bigskip\endinsert\global\advance\figno by1}
\def\endinsert{}

\begin{document}



\title{\large{\bf An Effective Field Theory Approach to Color
Superconductivity at High Quark Density}}

\author{ 
Nick Evans\thanks{nevans@budoe.bu.edu}
 \\ {\small Department of Physics, 
Boston University, Boston, MA 02215.} \\ \\
Stephen D.H.~Hsu\thanks{hsu@duende.uoregon.edu}
 \\ {\small Department of Physics, 
University of Oregon, Eugene OR 97403-5203.} \\ \\
Myckola Schwetz\thanks{myckola@baobab.rutgers.edu}
 \\ {\small Department of Physics and Astronomy, 
Rutgers University, 
Piscataway NJ 08855-0849.}\\ \\
}

\date{ August 1998 }

\maketitle

\begin{picture}(0,0)(0,0)
\put(350,386){OITS-657}
\put(350,374){RU-98-37}
\put(350,362){BUHEP-98-22}
\end{picture}
\vspace{-24pt}

\begin{abstract}
We investigate the Fermi Liquid theory of high density QCD. Using 
the renormalization group (RG), we determine the behavior of effective 
fermion interactions near the Fermi surface. At sufficiently high 
densities the matching between the Fermi Liquid theory and QCD can 
be accomplished in perturbation theory and a reliable calculation 
of Cooper pair formation performed, modulo the existence of a 
magnetic screening mass for the gluon. The presence of a chemical 
potential leads to different RG flows for sub-components of the 
effective interactions which would ordinarily be linked by Lorentz 
invariance. We also study instanton-induced operators and find that 
near the Fermi surface they are likely to be subdominant relative to 
operators induced by perturbative gluon exchange. We discuss the 
implications of our results for the phase structure of QCD at finite 
baryon density and temperature.

\end{abstract}

\newpage

\section{Introduction}

In this paper we investigate the dynamics of quark excitations
near the Fermi surface of high density QCD. Asymptotic freedom,
as well as explicit calculations \cite{FM}, suggest that the
high density limit is controlled by a small coupling: $\alpha_s$
evaluated at the length scale of typical quark separation. However,
it is well known that even very weak interactions can lead to 
interesting behavior in the presence of a Fermi surface  \cite{BCSrev}.
In particular, pair formation leading to a mass gap, superfluidity and
superconductivity seem generic whenever an attractive interaction, no
matter how weak, is present. 
One way to understand this is through the Fermi Liquid theory originally
developed by Landau, and recently clarified using the renormalization
group (RG) \cite{FLRG}. We review the RG method  in section 2
below within the context of relativistic theories. 
Our paper can be regarded as a first exploration of this
method applied to ultra-relativistic, non-Abelian gauge 
theories. Here we simply note that Fermi Liquid theory can be regarded
as an effective field theory of fermionic excitations, in which all 
interactions turn out to be irrelevant with the exception of four fermion
vertices with the kinematic structure associated with Cooper pairing 
(i.e. -- two quarks scattering with equal and opposite spatial momenta).
The renormalization group flows of these special operators therefore
determine the characteristics of the model.

At sufficiently high density we expect that the matching between the
fermionic effective theory and QCD can be accomplished in perturbation
theory (we discuss the details of this matching in section 3). 
The four fermion operators which are induced at leading order
are simply those from one gluon exchange with an IR cut off provided by
the gluonic screening masses. The gap equations resulting
from one gluon exchange were studied in some detail by Bailin and Love
\cite{BL}. As we will discuss
in section 4, their results are qualitatively similar 
to ours in the high density
regime, with the attractive $\bar{3}$ interaction 
leading to $0^+$ or $0^-$ diquark condensation and
color superconductivity. The RG approach provides an intuitive
understanding of this result since it is the attractive $\bar{3}$
interaction between two left (right) handed quarks that flows first
to a Landau pole as we approach the Fermi surface.

At somewhat lower densities, one might expect that non-perturbative interactions,
such as those mediated by instantons, play a role in the Fermi Liquid theory.
Recent attention \cite{instanton} has focused on the effects of instanton-inspired
fermionic models of QCD, although most of the analysis has been in the mean-field
approximation, with quantum corrections neglected. In section 5 we study the combined
RG flow of instanton and one-gluon induced vertices. We find that the one-gluon vertices
grow much faster as the Fermi surface is approached. Therefore, unless 
the instanton vertices are larger than the one-gluon vertices at the matching scale 
(which seems implausible), the dynamics is probably controlled by one-gluon effects. 
This suggests that the high density phase
identified above will persist to the (assumed) boundary with the low density phase which
exhibits chiral symmetry breaking.

Our paper is organized as follow. In section 2 we discuss the RG description of Fermi
Liquid theory for relativistic systems. In section 3 we describe the matching conditions
between QCD and our effective theory. In section 4 we derive and solve the RG equations
governing the flow of gluonic four fermion operators, obtaining an expression for
the condensate size as a function of quark density. In section 5 we study the combined RG
equations for the instanton and gluonic operators, and find that the gluonic operators
are dominant near the Fermi surface. In section 6 we discuss our results and their
implications for QCD at high densities and temperatures.

\section{Fermi Liquids from the Renormalization Group}

We wish to study QCD at finite density, which is described
by the following lagrangian with chemical potential $\mu$
\beq \label{lagrangian}
{\cal L } = - \frac{1}{4}F^{a \mu \nu} F^a_{\mu \nu} ~+~ 
\bar{\psi}_i ( i \DS + \mu  \gamma_0 ) \psi_i~~.
\eeq
We make a guess as to the form of the
effective theory close to the Fermi surface; the obvious guess based on
the dynamics of non-relativistic systems is that the theory is one of
weakly interacting fermions: these are the dressed ``quasi-particles''
of solid state physics language. We will henceforth refer to these effective 
degrees of freedom as ``quarks'', with the understanding that they may
in fact be related to the bare quarks in a complicated way.
Rather than treating the gluons as
propagating degrees of freedom we will integrate them out leaving a
potentially infinite sum over local, higher dimension fermion operators. 
The locality of these operators requires that gluons be screened at 
long-distances, presumably by effective electric and magnetic mass terms 
induced by the medium. 
We will address the matching between the high scale QCD dynamics and the
four Fermi theory in section 3 but for the moment we assume the matching
can be performed and proceed.

The Fermi surface in (\ref{lagrangian}) naturally breaks the O(3,1)
invariance of space-time to O(3) and furthermore picks out momenta of
order $\mu$. It is therefore natural to study the theory as we approach
the Fermi surface in a Wilsonian sense. We parameterize four momenta in the
following fashion
\beq 
p^{\mu} = (p_0, \vec{p}) = (k_0, \vec{k} + \vec{l})
\eeq
where $\vec{k}$ lies on the Fermi surface and $\vec{l}$ is perpendicular
to it. 
We study the Wilsonian effective theory of modes near the Fermi surface,
with energy and momenta  
\beq
\label{FSE}
|k_0|,|\vec{l}| ~<~ \Lambda~~~~,~~~~ \Lambda \rightarrow 0~~~.
\eeq  
While this type of RG scaling is somewhat
unfamiliar, it actually corresponds to thinning out fermionic degrees
of freedom according to their eigenvalues under the operator
$i \partial\!\!\!/  + \mu \gamma_0$. 
It is easy to see that eigenspinors of this operator with
eigenvalues $~\lambda_n:~ | \lambda_n | < \Lambda~$ correspond to states satisfying 
(\ref{FSE}). Consider an eigenspinor of the form
\beq
\psi_p ~=~ u(E,\vec{p})~ e^{ i( p \cdot x - p_0 t)}~~~,
\eeq
where $u(E,\vec{p})$ satisfies the usual momentum-space Dirac equation with
$E = |\vec{p}| = |\vec{k} + \vec{l}|= \mu ~\pm~ {\cal O} (\Lambda)$, and 
$p_0 = {\cal O} ( \Lambda )$. Then by direct substitution
we see that $\psi_p$ satisfies 
\beq
( i \partial\!\!\!/  + \mu \gamma_0 )~ \psi_k ~=~ {\cal O} ( \Lambda )~~~.
\eeq
Thus, the RG flow towards the Fermi surface just corresponds to taking 
the cutoff on eigenvalues of $i \partial\!\!\!/  + \mu \gamma_0$ to zero.

Which interactions are relevant operators in this limit? 
For our guess to make sense the kinetic term  
for the fermions must be invariant  when we
scale energies and momenta,
$k_0 \rightarrow s k_0$ (or, $t \rightarrow t/s$), 
and $\vec{l} \rightarrow s \vec{l}$,
 with $s < 1$. We must be careful though to
satisfy all the global symmetries of the theory. In particular, there is
a spurious symmetry of (\ref{lagrangian}) in which we treat $\mu$ as
the temporal component of a fictitious $U(1)_B$ gauge boson. 
In other words, the combined transformations 
$\psi \rightarrow e^{i \theta t}~ \psi$ and $\mu \rightarrow \mu - \theta$ 
leave the lagrangian (\ref{lagrangian}) invariant. From this, we deduce
that derivatives acting on $\psi$ may only enter the effective theory as
$i \partial\!\!\!/  + \mu \gamma_0$. This requires the kinetic term of our
effective theory to be of the form
\beq
\label{KE}
{\cal S}_{eff} = \int dt~ d^3p ~ \bar{\psi}^a_i \,
\Big( (i \partial_t   + \mu) \gamma_0 ~-~ \vec{p}\cdot \vec{\gamma}
\Big)\, \psi^a_i
\eeq
In the Wilsonian RG scaling, we eliminate all modes with
energy and momenta  $|k_0|,|l| > \Lambda$, where $\Lambda$ is our cutoff. 
As discussed above, on the remaining degrees of freedom the operator 
$(i \partial_t  + \mu) \gamma_0 - \vec{p}\cdot \vec{\gamma} 
\sim {\cal O}(\Lambda)$ and therefore scales like $s$.
We deduce that for (\ref{KE}) to remain invariant,  
$\psi$ must scale as $s^{-1/2}$. The term representing the
Fermi surface scales as $s^{-1}$ indicating that it is a relevant operator
and that hence the existence of the surface is natural.
Now consider the four fermion operator
\beq
\label{ffop}
i G \int dt\, d^3\vec{p_1}\,d^3\vec{p_2}\,d^3\vec{p_3}\,d^3\vec{p_4}  ~
\bar{\psi}_k \Gamma^\mu_{ki} \psi_i ~ \bar{\psi}_l \Gamma_{\mu lj} \psi_j
\eeq
where $\Gamma$ contains any gamma matrix and flavor or color structure. 
Naively, (\ref{ffop}) is irrelevant since it scales as $s$
if we assume we have $\vec{l}$ close to zero and
hence
\beq
\delta^3(\vec{p_1} + \vec{p_2} - \vec{p_3} - \vec{p_4} ) \simeq 
 \delta^3(\vec{k_1} + \vec{k_2} - \vec{k_3} - \vec{k_4} )~~~,
\eeq
which does not scale. Higher dimension terms with extra powers of the
fields or terms with additional
derivatives are clearly irrelevant operators as well. 
(Again, because of the spurious symmetry higher derivative terms
must enter in combination with $\mu$, and hence scale like $s$.)
The only operators
that survive are those satisfying the kinematic constraint 
$\vec{p}_1 \simeq - \vec{p}_2$, so the
delta function becomes
\beq
\delta(\vec{l_1} + \vec{l_2} - \vec{l_3} - \vec{l_4})
\eeq
and scales as $s^{-1}$ ($\int dl \delta(l)=1$). 
The resulting four fermion operator is marginal.

The only interactions in the effective theory are therefore four fermion 
operators at a special kinematic point, and the model is exactly soluble
by resummation. 
The only diagrams allowed by the
momentum structure are the bubble diagrams 
found at large N in the familiar O(N) model. The interaction generates
cooper pair formation through the gap equation (see Bailin and
Love in \cite{BL})
\beq
\Delta = iG \int_0^{\Lambda_{UV}} {d^4p \over (2 \pi)^4 } \Gamma^\mu C(p)
\Gamma_\mu ~~~,
\eeq
where G is the coupling, $\Gamma^\mu$ any associated Dirac 
structure and $C(k)$ a $4\times 4$ off diagonal component of the
$8\times 8$ propagator associated with the fermion vector $(\psi,
\psi^C)$ 
\beq
C(p) = {1 \over ( p\!\!\!/  - \mu \gamma_0)} ~ \Delta ~ {1 \over 
[ \tilde{\Delta} ( p\!\!\!/  - \mu
\gamma_0)^{-1} \Delta - ( p\!\!\!/ + \mu \gamma_0)] }
\eeq
where $\tilde{\Delta} = \gamma_0 \Delta^\dagger \gamma_0$.
The gap integral would diverge logarithmically as $p_0 \rightarrow 0$ and
$|\vec{p}|^2 \rightarrow \mu^2$, were it not for the
factors of $\Delta$ in the denominator of $C(p)$ which provide
an IR cutoff. Ignoring the Lorentz structure of the integral for the
moment, we have
\beq
\Delta  \sim  -i G \int^{\Lambda_{UV}}_\Delta {d^4p \over (2 \pi)^4 } {1
  \over (p\!\!\!/ - \mu \gamma_0)} \Delta {1 \over (p\!\!\!/  + \mu \gamma_0)} 
~~,
\eeq
or,
\begin{eqnarray}
\Delta &\sim &
{ - iG \Delta \over 4}\int^{\Lambda_{UV}}_\Delta {d^2k~ dl~ dk_0 \over (2
  \pi)^4} {1 \over (k_0 +l-i\epsilon) (k_0-l+i\epsilon)}\\
&\sim &
 {G N \Delta \over 4} \ln \left({ \Lambda_{UV} \over \Delta} \right) ~~~.
\end{eqnarray}
Here $N = \int d^2k / (2 \pi)^3 = \mu^2/ 2 \pi^2$ is the density of
states at the Fermi surface in the lowest order approximation.
The Cooper pair condensate is of the form
\beq
\Delta = \Lambda_{UV}~ e^{- {c  \over G}}~~,
\eeq
where $c$ is a constant.

Thus we obtain the result (well known in the context of non-relativistic
superconductivity) that
any weakly attractive operator of this form generates cooper pair
formation. A repulsive interaction ($G<0$) only has the consistent
solution $\Delta =0$. This conclusion in the relativistic case depends
on the precise Lorentz matrix structure of the interaction and the
condensate and is the subject of study below. 
Although the condensate appears to be formed due to weak interactions
this is a misleading consequence of the exact solubility of the gap equation. 
In fact as we will show in section 4 the couplings
become strong  as the effective theory approaches the Fermi
surface, and from this point of view it is the strong coupling physics that 
generates the condensate. By analogy, 
if one could solve the Swinger-Dyson equations
of zero density QCD at short distance, where the coupling is small, 
one might incorrectly conclude that the usual
chiral condensate results from weak coupling effects, while it is in
fact due to strong IR effects. 

To determine the behavior of QCD at high density we must first perform 
the matching between QCD and the four Fermi theory (section 3). 
We will see that several types of four fermion interaction are generated. 
In section 4 we will perform a RG analysis of how
those couplings run to the Fermi surface to identify the dominant
attractive interaction.

\section{Matching to QCD}

At very high density, $\mu \gg \Lambda_{QCD}$, we expect the
four quark coupling to be dominated by one gluon exchange due to 
asymptotic freedom. By taking $\mu$ sufficiently large $g(\mu)$ can be
made arbitrarily small and one loop effects are then 
negligible. At tree level QCD generates several four fermion
vertices. A pair of quarks in the 6 representation of $SU(3)$ 
can scatter to a 6, and quarks in the $\bar{3}$ can scatter to 
the $\bar{3}$. However, the 6 and $\bar{3}$
can not mix. In addition because the introduction of $\mu$ breaks the
$O(3,1)$ invariance of the theory we must treat the temporal and spatial
components of the couplings independently.  
We thus have four interactions in the four fermi theory:
\beq \label{couple}
\begin{array}{c}
 iG_{6}^{0}  ~
 (\delta_{ca} \delta_{db} + \delta_{cb} \delta_{da}) 
\bar{\psi}_c \gamma_0 \psi_a \bar{\psi}_d \gamma_0 \psi_b  \\
\\
 iG_{6}^i ~ 
 (\delta_{ca} \delta_{db} + \delta_{cb} \delta_{da}) 
\bar{\psi}_c \gamma_i \psi_a \bar{\psi}_d \gamma_i \psi_b  \\
\\
 i G_{\bar{3}}^{0}  ~ 
  (\delta_{ca} \delta_{db} - \delta_{cb} \delta_{da}) ~ 
\bar{\psi}_c \gamma_0 \psi_a \bar{\psi}_d \gamma_0 \psi_b  \\
\\
 i G_{\bar{3}}^{i} ~ 
(\delta_{ca} \delta_{db} - \delta_{cb} \delta_{da}) ~ 
\bar{\psi}_c \gamma_i \psi_a \bar{\psi}_d \gamma_i \psi_b  \\
\end{array}
\eeq
where we have used the decomposition
\beq 
(T^a)_{ca} (T^a)_{db} = {1 \over 6} (\delta_{ca} \delta_{db} +
\delta_{cb} \delta_{da}) - {1 \over 3}  (\delta_{ca} \delta_{db} -
\delta_{cb} \delta_{da}) ~~.
\eeq
Henceforth we will include signs from the contraction of spacelike
$\gamma_i$ matrices in the coupling constants. 
The one-gluon approximation yields the following matching
conditions:
\beq \begin{array}{ccc}
G^0_6 = {1 \over 6} {4\pi \alpha_s \over k^2}
 & \hspace{1cm} &  G^0_{\bar{3}}  =    - {1 \over 3}
{4\pi \alpha_s \over k^2} \\
G^i_6 =  -{1 \over 6} {4\pi \alpha_s \over k^2} &  & 
G^i_{\bar{3}} =  {1 \over 3}
{4\pi \alpha_s \over k^2} ~~,
\end{array}
\eeq
where $k$ is the four-momentum in the gluon propagator. 

However, this is too naive and leads to non-local interactions
dominated by arbitrarily soft gluon exchange ($k \rightarrow 0$), 
with all the consequences of asymptotic freedom.
It is reasonable to assume that the non-zero quark density screens
long-range gluon exhange, leading to an effective gluon ``mass''.
At one loop order, a chemical potential leads to Debye screening of
the temporal component of the gluon that can be calculated
perturbatively \cite{FM}. The details are not important here, but
the result is of the form
\beq
M_0^2 \sim g^2 \mu^2~~~.
\eeq
It is generally believed that
the spatial or magnetic components of the gluons acquire masses
only non-perturbatively. If the finite density behavior is similar
to the finite temperature case, one would expect
\beq
M_i^2 \sim g^4 \mu^2~~.
\eeq
In this paper we merely assume the existence of both a magnetic
and electric screening mass, with the electric mass somewhat larger.
Our results will be qualitatively independent of the precise values
of these masses, which provide a cutoff on infrared effects.
In the four fermion theory
we therefore introduce a form factor
\beq
\label{FF}
{ i g^2 \over (k^2 - M_{0,i}^2)}  ~=~ { -i g^2 \over M^2_{0,i} }
\left( {1 \over 1 + k^2  / M^2_{0,i} } \right) ~=~  { -i g^2 \over M^2_{0,i} }
 \left( 1 -   {k^2 \over M_{0,i}^2} + ~...~ \right)~~~.
\eeq
Since $k_0 \rightarrow 0$ as we approach the Fermi surface,
$k^2 < 0$, and the largest interactions come from
scatterings with small momentum transfer.
In fact, our previous arguments concerning the 
spurious $U(1)_B$ symmetry imply that 
the higher order terms in this expansion are suppressed (correspond to
irrelevant operators) near the Fermi surface. Hence in the $s \rightarrow 0$
limit we need only retain interactions with constant form factor (no derivatives) in
our RG analysis. In the BCS analysis \cite{BCSrev} a constant potential is also
used, and this is usually regarded as a crude approximation. However, the symmetry
argument implies that it is accurate near the Fermi surface. Note that the
symmetry argument does {\it not} imply that the $k^2 / M^2$ terms can be neglected
at the matching scale. Indeed, one can easily see that neglecting the higher order
terms on the r.h.s. of (\ref{FF}) leads to a large overestimate of the strength of
the interaction. Note that we
have performed the matching in Feynman gauge. In the one loop approximation 
results such as the location of the Landau pole are gauge invariant, so we can
make any choice of gauge. However, some gauge dependence may arise from the
gluon masses, which we have inserted by hand.

In order to obtain an accurate matching of a four fermion operator
with constant form factor to the one-gluon interaction, we average the
momentum dependent function in (\ref{FF}) over all incoming and outgoing
quark momenta, with both quarks near the Fermi surface. This is equivalent
to projecting out the zero angular momentum component of the operator. It is
possible to separate any operator into eigenstates of angular momentum,
which do not mix under the RG flow. Since we expect that the eventual condensate
will be in the zero angular momentum channel (see \cite{BL} for a careful
justification of this in the case of the one-gluon exchange kernel), we do
not lose any information by only retaining this component.
Let $\vec{p}$ and $\vec{q}$ denote the incoming and outgoing momenta of a quark line, and
\beq
k^2 = (p - q)^2 = p^2 +q^2 - 2p \cdot q ~~.
\eeq
For $\vec{p}$ and $\vec{q}$ near the Fermi surface, and $\theta$ the angle
between $\vec{p}$ and $\vec{q}$, this becomes
\beq
k^2 = - 2 \mu^2 ( 1 - \cos \theta )~~.
\eeq
Averaging the form factor (\ref{FF}) over all relative orientations
of $\vec{p}$ and $\vec{q}$, we obtain
\beqa
\label{average}
{1 \over 4 \pi \mu^2} \int d^2k ~{ 1 \over k^2 - M^2} &=&
{1 \over 4 \mu^2} \int_{-1}^{+1} dz ~{  1 \over 1 - z + M^2/2 \mu^2 } \nonumber  \\
&=& {1 \over 4 \mu^2} \ln \left[ { 4 \mu^2 \over M^2} \right]~~~.
\eeqa
We see that in this approximation the matching conditions depend
only logarithmically on the gluon masses $M_{0,i}$. A more sophisticated
scheme for estimating the strength of the constant form factor of the
four fermion interaction might yield a slightly different relation
than (\ref{average}), but we expect it to still depend only weakly
on the screening masses. The logarithm
in (\ref{average}) is of order $\ln [g^{-2}]$ or $\ln [g^{-4}]$, and is
somewhat larger than one.

Our matching conditions are then:
\beq \begin{array}{ccc}
G^0_6 = -{1 \over 6} {\pi \alpha_s( \mu ) \over \mu^2} \ln \left[ { 4 \mu^2 \over M^2} \right]
 & \hspace{1cm} &  G^0_{\bar{3}}  =     {1 \over 3}
{\pi \alpha_s( \mu ) \over \mu^2} \ln \left[ { 4 \mu^2 \over M^2} \right] \\
G^i_6 =  {1 \over 6} {\pi \alpha_s( \mu ) \over \mu^2} \ln \left[ { 4 \mu^2 \over M^2} \right] &  & 
G^i_{\bar{3}} =  - {1 \over 3}
{\pi \alpha_s( \mu ) \over \mu^2} \ln \left[ { 4 \mu^2 \over M^2} \right] ~~~.
\end{array}
\eeq

\noindent The scale of this matching is of order $\mu$, and corresponds to the typical momentum
transfer in the scattering of two quarks on the Fermi surface.

\vspace{1cm}

$\left. \right.$  \hspace{-1cm}\ifig\prtbdiag{}
{\epsfxsize8truecm\epsfbox{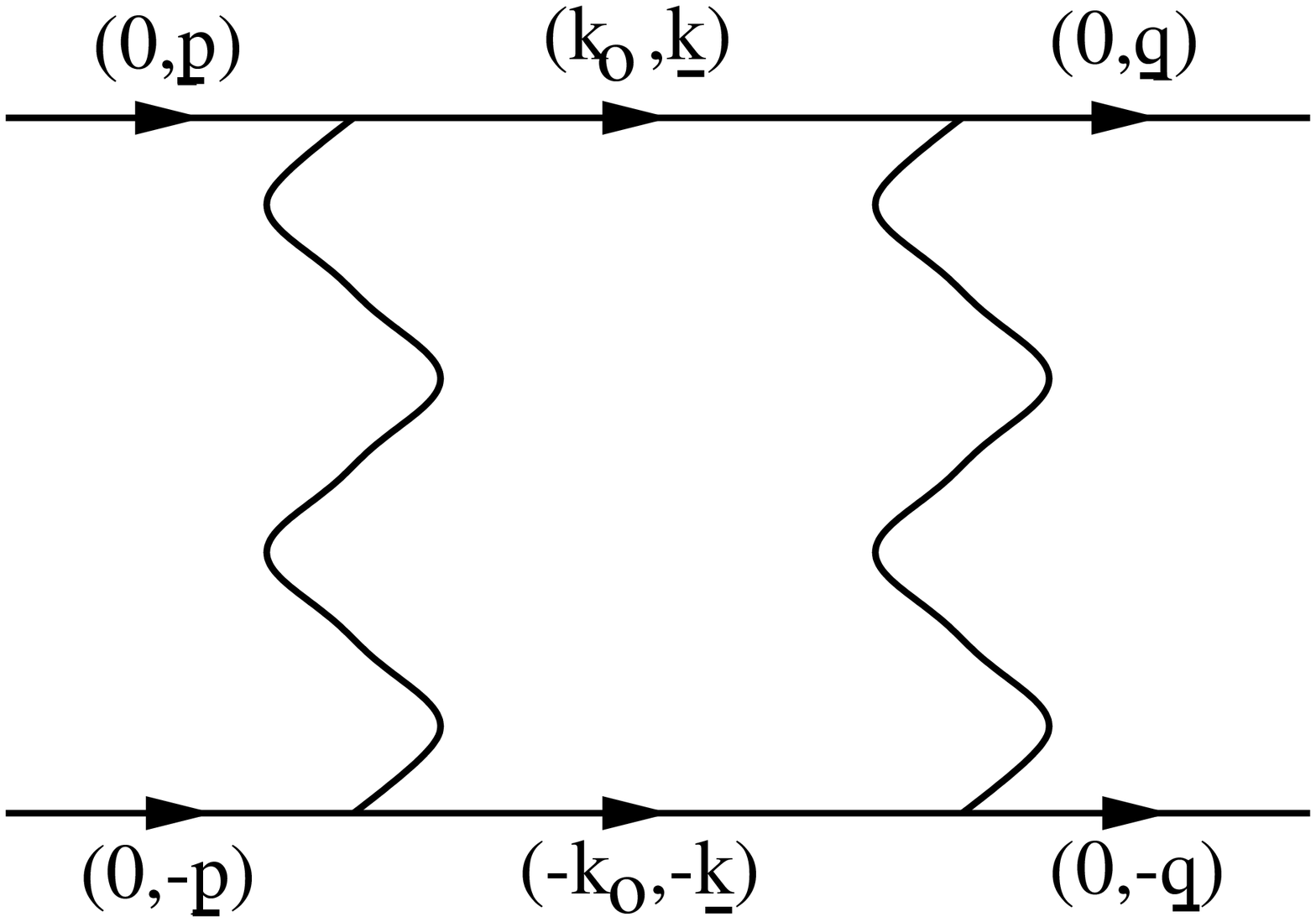}} 

\begin{center}
Figure 1: One loop diagram from gluon exchange which drives the RG flow of
effective four fermion interactions.
\end{center}

\section{One Gluon RG Flow}

The Fermi liquid displays interesting behavior such as superconductivity
because in the effective theory the marginal four fermion interactions
flow to strong coupling as they approach the Fermi surface. To display
the basic behavior, consider a theory with just the simple interaction
\beq
 i G \bar{\psi} \psi \bar{\psi} \psi~~~.
\eeq
This interaction leads to a one loop graph which
is the analog of figure 1 with the gluon lines
contracted. (Note that because of the restricted momentum structure
of the vertices in the effective theory the diagram with crossed gluon
lines has no counterpart and does not play a role in the RG flow.)
It contributes a logarithmic divergence to the 
running in the presence of a Fermi surface:
\beq
- G^2 \int {d^4p \over (2 \pi)^4} \left[{i \over p^\mu\gamma_\mu + 
\mu \gamma_0 - i \epsilon} \right]_{ik} \left[
 {i \over -p^{\nu} \gamma_\nu + \mu \gamma_0 - i \epsilon} \right]_{jl}
\eeq
Performing the gamma matrix algebra and taking
the limit $k_0, |\vec{l}| \rightarrow 0, 
|\vec{k}|^2 \rightarrow \mu^2$, this
becomes
\beq
\label{gamma}
G^2  \left[  - (\gamma_0)_{ij} (\gamma_0)_{kl} + {1 \over 3}
(\gamma_\alpha )_{ij} (\gamma_\alpha)_{kl}\right]  I
\eeq
where the log divergent part of the integral I is given by
\begin{eqnarray}
I & = &  {1\over 4} 
 \int {dk_0~ d^2k~ dl \over (2 \pi)^4} {1 \over (k_0+l-i \epsilon)(k_0-l
+ i \epsilon)} \nonumber \\
 & \simeq & {i \over 4 } N  \ln \left( {\Lambda_{IR} \over \Lambda_{UV}}
\right)~~~.
\end{eqnarray} 
$N = \int d^2k / (2 \pi)^3 = \mu^2/ 2 \pi^2$ is the density of
states at the Fermi surface in the lowest order approximation.
Here we have moved from an effective theory with cutoff $\Lambda_{UV}$
($k_0 , |\vec{l}| <  \Lambda_{UV}$), 
to a new effective theory with cutoff $\Lambda_{IR}$.
As we approach the Fermi surface ($\Lambda_{IR}
\rightarrow 0$), the coupling $G$ runs logarithmically. 

In this fashion we can calculate the one loop beta functions for the
vertices in (\ref{couple}) as a function of $t=\ln(\Lambda_{IR}/\Lambda_{UV})$. 
It is most convenient to perform the calculation (in the massless limit)
using two component Weyl spinors. The gluon exchange conserves helicity
and flavor quantum numbers. 
From figure 1 and (\ref{gamma}) 
it can be seen that the graphs, in two component notation, involve 
the contractions of 6 sigma matrices (associated with the gauge vertices
and resulting from the loop integral). The following 
$\sigma$ matrix identities are useful
\begin{eqnarray}
(\sigma^i)_{ab}(\sigma^j)_{bc}(\sigma^k)_{cd}(\sigma^i)_{xy}(\sigma^j)_{yz}
(\sigma^k)_{zw} & = & 7  (\sigma^i)_{ad}(\sigma^i)_{xw} - 6
(\sigma^0)_{ad} (\sigma^0)_{xw}\\
&& \nonumber \\
(\sigma^i)_{ab}(\sigma^j)_{bc}(\sigma^0)_{cd}(\sigma^i)_{xy}(\sigma^j)_{yz}
(\sigma^0)_{zw} & = & -2  (\sigma^i)_{ad}(\sigma^i)_{xw} + 3 
(\sigma^0)_{ad} (\sigma^0)_{xw}\\
&& \nonumber \\
(\sigma^i)_{ab}(\sigma^0)_{bc}(\sigma^0)_{cd}(\sigma^i)_{xy}(\sigma^0)_{yz}
(\sigma^0)_{zw} & = &   (\sigma^i)_{ad}(\sigma^i)_{xw} \\
&& \nonumber \\
(\sigma^0)_{ab}(\sigma^0)_{bc}(\sigma^0)_{cd}(\sigma^0)_{xy}(\sigma^0)_{yz}
(\sigma^0)_{zw} & = &  (\sigma^0)_{ad}(\sigma^0)_{xw}
\end{eqnarray}
The gamma matrix algebra at one loop distinguishes the
couplings between two left (right) handed quarks and those between  left and
right handed quarks. This is a result of the divergence in the diagram of
figure 1. Without a chemical potential this diagram is not divergent for the 
restricted shell of momenta under consideration, and this discrepancy 
between the two interactions is not seen. The RG eqns are
\begin{eqnarray}
{d G^{0}_{a LL} \over dt} & = & {N \over 2} \left( -
(G^{0}_{a LL})^2 + 2 G^{0}_{a LL}\, G^{i}_{a LL} - 5 
(G^{i}_{a LL})^2 \right) \\
&&\nonumber \\
{d G^{i}_{a LL} \over dt} & = & { N \over 2} \left( {1
  \over 3}
(G^{0}_{a LL})^2 - {10 \over 3}
 G^{0}_{a LL}\, G^{i}_{a LL} + {13 \over 3}  
(G^{i}_{a LL})^2 \right) \\
&&\nonumber \\
{d G^{0}_{a LR} \over dt} & = & {N \over 2} \left( -
(G^{0}_{a LR})^2 + 2
 G^{0}_{a LR}\, G^{i}_{a LR} -  
(G^{\i}_{a LR})^2 \right) \\
&&\nonumber \\
{d G^{i}_{a LR} \over dt} & = & {N \over 2} \left( 
{1 \over 3} (G^{0}_{a LR})^2 - {2 \over 3} 
 G^{0}_{a LR}\, G^{i}_{a LR} + {1 \over 3}  
(G^{i}_{a LR})^2 \right) ~~~.
\end{eqnarray}
We suppress the RR and RL equations since they are respectively identical
to the LL and LR equations already displayed.
The $a$ subscript corresponds to either  6 or $\bar{3}$; note that
again loop effects do not mix the two channels. 

These equations diagonalize to
\begin{eqnarray}
{d (G^{0}_{a LL} + G^i_{a LL}) \over dt} & = & -  {1 \over 3} N 
( G^{0}_{a LL} + G^i_{a LL})^2 \\
&&\nonumber \\
{d (G^{0}_{a LL} - 3G^{i}_{a LL}) \over dt} & = & -  N 
(G^{0}_{a LL} - 3G^{i}_{a LL})^2 \\
&&\nonumber \\
{d (G^{0}_{a LR} + 3 G^i_{a LR}) \over dt} & = & 0 \\
&&\nonumber \\
{d (G^{0}_{a LR} - G^{i}_{a LR}) \over dt} & = & - {2 \over 3} N 
(G^{0}_{a LR} - G^{i}_{a LR})^2
\end{eqnarray}
Taking $G^0 = G(0)$ and $G^i = -x G(0)$ as the boundary
condition at $t=0$, we obtain the solutions
\begin{eqnarray}
G^0_{aLL}(t) & = & {1 \over 4} \left( { 3 G(0)(1-x) \over 1 + {N \over 3}
  G(0) (1-x) t } + {G(0) (1+3x) \over 1 + N G(0) (1+3x) t }
  \right)\\
&&\nonumber \\
G^i_{aLL}(t) & = & {1 \over 4} \left(  {  G(0)(1-x) \over 1 + {N \over 3}
  G(0) (1-x) t } -  {G(0) (1+3x) \over 1 + N G(0) (1+3x) t}
  \right)\\
&&\nonumber \\
G^0_{aLR}(t) & = & {1 \over 4} \left( G(0) (1-3x) +  {3G(0) (1+x) 
\over 1 + {2N \over 3} G(0) (1+x) t }
  \right)\\
&&\nonumber \\
G^i_{aLR}(t) & = & {1 \over 4} \left( G(0) (1-3x) -  {G(0) (1+x) 
\over 1 + {2N \over 3} G(0) (1+x) t} \right)
\end{eqnarray}

We may now consider the boundary conditions from section 3. We expect
$x > 1$, since the Debye mass appears at lowest order in perturbation
theory and the magnetic mass is a higher-order (non-perturbative) effect. 
The Fermi surface lies at $t \rightarrow - \infty$.
In the the $\bar{3}$ channel $G_{\bar{3}}(0)$ is positive and the LL (and RR)
couplings have a Landau pole at 
\beq
\label{Lpole}
t = - 1/NG_{\bar{3}}(0)(1+3x)~~~,
\eeq
satisfying the
asymptotic relation $G^0_{\bar{3}LL} = - G^i_{\bar{3}LL}$. 
The LR (and RL) couplings also have a
Landau pole at $t = - 3/2NG_{\bar{3}}(0)(1+x)$ but this lies closer to
the Fermi surface than the LL coupling pole for all $x > 1$. 
As we approach the pole in the LL (and RR) channel 
the LR (and RL) coupling becomes negligible in comparison.

In the $6$ channel $G_{6}(0)$ is negative and smaller than
$G_{\bar{3}}(0)$. The interaction remains repulsive and the
LL (and RR) couplings have a Landau pole at $t = - 1/3NG_{6}(0)(x-1)$. 
The LR couplings flow to constant values as the Fermi surface 
is approached. 

We conclude that the Cooper pairing dynamics is 
dominated by the attractive $\bar{3}$ interaction
between two left (right) handed quarks. 
The condensate that forms is therefore
antisymmetric in color. The LL and RR channels correspond to the 
antisymmetric spin 0 state and the condensate is either
$\psi^T C \psi$ or $\psi^T C \gamma_5 \psi$ .
The one gluon exchange diagrams do not distinguish between 
the scalar and pseudo-scalar condensates, because the two are related by 
the $U(1)_A$
symmetry which at the level of one gluon exchange is a good 
symmetry.
The dynamics is flavor blind but the flavor of the
condensate is determined by Fermi statistics. It must be in the
anti-symmetric singlet of flavor in the case of two light quarks.
In the case of three light quarks there is the possibility of
color-flavor locking \cite{CFL}. 
In reality we expect the large strange quark mass to displace its Fermi 
surface relative to that of the up and down quarks. This mismatch of Fermi
momenta disfavors mixed strange -- nonstrange condensates, 
except possibly at lower densities where
the condensate might be quite large. The mismatch in Fermi momenta is
roughly
\beq
\Delta p_F ~=~ p_F - p_{Fs} ~=~ \mu - \sqrt{\mu^2 - m_s^2} ~~~, 
\eeq
where $\mu$ is the common chemical potential for the u,d and s
quarks. This will be relevant when the condensates are of order or less than
$\Delta p_F$. For 
$\mu = 300$ MeV and $m_s = 150$ MeV, $\Delta p_F \simeq 40$ MeV. 
In the absence of mixed condensates the strange quarks will condense
on their own in the $\bar{3}$ channel which is  antisymmetric in color. 
There is only a flavor symmetric state available in this case.
Thus the LL (RR) interaction can 
only cause the condensation of the antisymmetric spin 0 and angular momentum
$L=1$ state which would be smaller than the $u$ and $d$
condensate. There is also potentially competition from the LR (RL)
couplings which have a Landau pole closer to the Fermi surface and could
give rise to condensation in the  symmetric spin 1 and $L=0$ state.
Again this condensate, if it formed, would be smaller than that in the $u$
and $d$ sector. 

We are now in a position to estimate the size of the actual
condensate, by computing the RG invariant scale (location of
the Landau pole (\ref{Lpole})) at which the $\bar{3}$, LL coupling 
diverges. This yields
\beq
\label{cond1}
\Delta ~=~ \Lambda_{UV}~ \exp \left( - ~{ 1 \over N G_{\bar{3}}(0)~ (1 + 3x)} \right)~~.
\eeq
Using the matching conditions from section 3, and the free-quark
approximation to the density of states $N$, we have
\beq
\label{cond2}
\Delta ~=~ \Lambda_{UV}~ \exp \left( - { 6 \pi \over 1 + 3x}~{1 \over \alpha_s}~~
\ln \left[ {4 \mu^2 \over M_i^2} \right]^{-1} \right)~~.
\eeq
Note that for densities just above the expected phase
boundary for chiral symmetry restoration, $\alpha_s \sim {\cal O} (1)$,
and (\ref{cond2}) predicts a relatively large condensate. As we increase
the density, the condensate decreases. In order to compute the rate of
decrease, we use the QCD beta function relation
\beq
\Lambda_{QCD} ~=~ \Lambda_{UV}~ \exp \left( - {2 \pi \over b_0} {1 \over 
\alpha_s ( \Lambda_{UV}) } \right),
\eeq
where $b_0 = 11 - 2N/3$ for N flavors. Rewriting (\ref{cond2}) in terms of
$\Lambda_{QCD}$, we find
\beq
\label{cond3}
{ \Delta \over \Lambda_{QCD} } ~=~ 
 \left( {\Lambda_{QCD} \over \Lambda_{UV} } \right)^{q} ~~~,
\eeq
where 
$$
q =  {3 \over 1 + 3x}~ {b_0 \over  \ln [ {4 \mu^2 \over M_i^2} ]} ~-~ 1~~~.   
$$
Recalling that $\Lambda_{UV} \sim  \mu$, we see that the condensate scale
$\Delta$ decreases rapidly with increasing $\mu$. Incidentally, equation
(\ref{cond3}) again shows that the system can naturally support 
$\Delta \sim \Lambda_{QCD}$, since at low density we expect 
$\Lambda_{UV} \sim \Lambda_{QCD}$. We have checked that our results in
(\ref{cond2}) and (\ref{cond3}) are qualitatively consistent with the
earlier gap equation calculations of \cite{BL} and \cite{instanton}.

Finally, we note that (\ref{cond2}) actually predicts that the
condensate will eventually increase at asymptotically high densities,
due to the logarithm in the exponent. It is not clear how reliable this
prediction is, as it depends sensitively on the matching (averaging) procedure
and on the precise form of the magnetic screening  mass $M_i$.

\vspace{1cm}

$\left. \right.$  \hspace{-1cm}\ifig\prtbdiag{}
{\epsfxsize8truecm\epsfbox{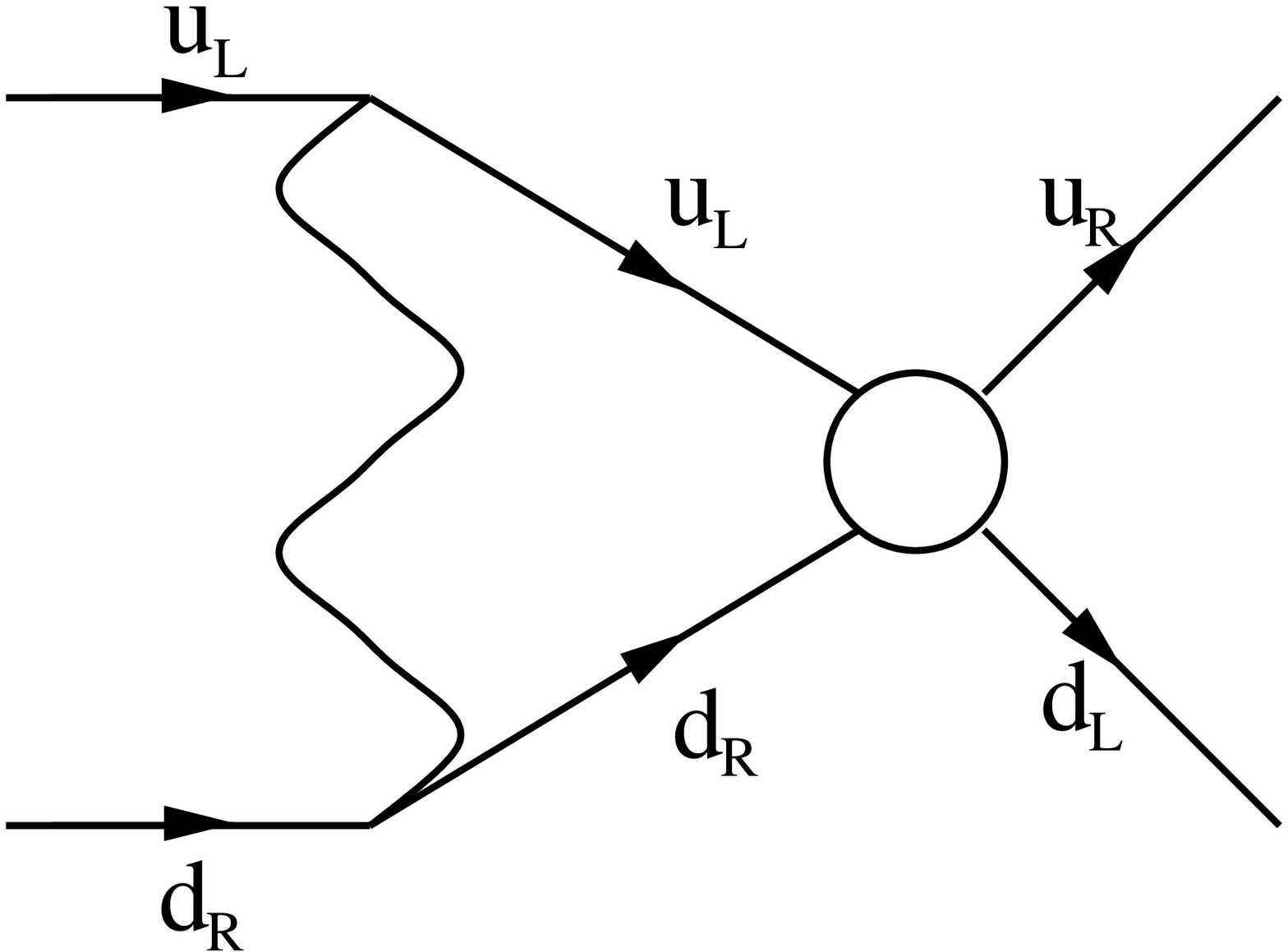}}  
\begin{center}
Figure 2: The gluonic correction to the instanton induced vertex.
\end{center}

\section{Instanton RG Flow}

Recently a number of groups \cite{instanton} have considered the possible role of
instanton induced interactions in color superconductivity, particularly at 
low densities. In this section we study the combined RG flow of instanton
and gluon--induced interactions. At sufficiently high density, instanton
effects are almost certainly negligible. As we lower the density, the
relative size of the instanton--induced operators at the matching scale
should grow relative to the gluon operators. At sufficiently low densities
we can imagine that the two types of interactions become roughly comparable
in strength. The RG equations will tell us which interactions dominate the
dynamics at the Fermi surface.

In the two-flavor case, which we analyse here, 
the instanton-induced operators are of the form
\beq
\kappa T^a T^a \left( \bar{u}_R u_L \bar{d}_R d_L - \bar{u}_R d_L
\bar{d}_R u_L \right)~~.
\eeq
We assume that at the matching scale the instanton-induced coupling 
$\kappa(0)$ is smaller than the gluon-induced coupling $G(0)$. If this
hierarchy persists as $t \rightarrow - \infty$, then it is self-consistent
to neglect loop diagrams with multiple instanton vertices. In this case
the RG flow of $\kappa$ is determined by the effective field theory
counterpart of the diagram in
fig 2, which depicts the instanton operator renormalized by gluon exchange. 
In the effective theory we compute the loop diagrams with one
gluon and one instanton type vertex.
We obtain the following one loop beta functions, where again the subscript
$a$ denotes either the $6$ or $\bar{3}$ channels.
\beq 
\label{IRGE}
{d \kappa_{a} \over d t} = N \kappa_a (- G^0_{aLL} + G^i_{aLL})~~~.
\eeq

Solving (\ref{IRGE}), we can directly compare the size of $\kappa_a$ to
$G^0_{\bar{3}LL}$ (the largest gluonic coupling). We find
\beq 
\label{ratio}
{\kappa_{a}(t) \over G^0_{\bar{3}LL}(t) } =
{ \kappa_{a}(0) \over G^0_{\bar{3}LL}(0)}{4 \over (1+3x) }~ \left( 1 +
NG(0)(1+3x)~t \right)^{1/2}~~~,
\eeq
which shows that as we approach the Fermi surface the instanton interaction
becomes more and more suppressed relative to the gluonic operator. 
When the gluonic $\bar{3}$ interaction reaches its Landau pole 
at $t = - 1 / N G(0) (1+3x)$, the instanton interaction is completely
negligible. This result implies that as long as the instanton interaction
is somewhat smaller than the gluon interaction at matching, the pattern
of condensation will be determined solely by gluonic effects.

The diagram in figure 2 also induces operators with different Lorentz 
structure from the bare vertices, as do graphs with multiple instanton
vertices. For example, an interaction of the form 
$(\bar{q}_R \gamma_0 \gamma_i q_L) (\bar{q}_R \gamma_0 \gamma_i q_L)$ 
is induced.
The ratio of this new coupling to $G^0_{\bar{3}LL}$ behaves similarly to 
(\ref{ratio}) and thus vanishes as one approaches the Fermi surface.
In general,  new vertices are always smaller than the bare couplings
if we take $\kappa(0)$ somewhat less than $G(0)$. 
However, if $\kappa(0)$ and $G(0)$ are of the same size, as is
possibly the case at lower densities, then the RG equations become
considerably more complicated and the instanton vertex may play a role
in Cooper pair formation.

\section{Discussion}

The renormalization group provides a powerful tool for the analysis
of QCD at large quark density. The highly constrained nature of the
effective field theory describing quarks near the Fermi surface allows
a detailed analysis of the relevant interactions. We find that the
operators induced by gluon exchange lead to Cooper pair (diquark)
condensation and color superconductivity in the $\bar{3}$, $0^\pm$ 
channels. The relevant order parameter for the two-flavor condensate is
\beq
\label{condensate}
\psi^T_{ia}(-p) ~C ~ \left( \cos( \theta ) + i \sin ( \theta )  
\gamma_5 \right) ~ \psi_{jb}(p) ~ 
\epsilon^{ij}~\epsilon^{abc} ~~~,
\eeq
where $C$ is the charge-conjugation matrix, 
(i,j) are flavor indices running from 1 to 2, and (a,b) are color indices.
$\theta = \pi / 2$ preserves parity, while other values break it spontaneously.
Since the different $\theta$ condensates are related by the $U(1)_A$ symmetry,
which is good to all orders in perturbation theory, the gluon interactions
cannot differentiate between the two possibilities, and it is up to interactions
like instantons to determine whether parity is violated \cite{instanton}.  
At high densities, where instanton effects are negligible, explicit quark masses
also violate the $U(1)_A$ symmetry, and we have checked that the parity conserving
vacuum is the true minimum. Nevertheless, at high density the entire family of 
parity violating and parity conserving 
vacua are nearly degenerate. This implies the existence of a light 
pseudo-Goldstone boson analogous to the $\eta'$, except that it 
corresponds to rotations in 
the diquark condensate rather than the usual quark condensate. This
particle is a color singlet, pseudo-scalar. 
Due to this excitation the mass gap in the physical
spectrum can be much smaller than the condensate scale $\Delta$.

Our calculations were performed assuming a sufficiently high density to
justify perturbative matching to QCD. Strictly speaking, they should only
be applied at densities somewhat larger than the critical density where
the chiral phase transition is expected to occur. Nevertheless, they yield
results which can be naively extrapolated to lower densities. Equation 
(\ref{cond3}) suggests that a condensate of order $\Lambda_{QCD}$ is possible.

We also studied instanton induced operators in the effective theory. 
We found that the RG flow of such operators is slower than that of 
the one gluon operators, and hence that the latter reach their Landau pole
first. This lends further confidence that the one gluon results can be
extrapolated to lower densities. However, it is still possible
that non-perturbative effects affect diquark condensation, depending on the
relative sizes of instanton and gluon operators at the matching scale.

Finally, we note that all of our calculations have been at zero temperature.
A nonzero temperature would smear out the Fermi surface, eventually eliminating
the condensates. Bailin and Love \cite{BL} studied the phase transition between 
the superconducting and normal states in the one gluon exchange approximation, 
and found a fluctuation-induced first order phase transition. Since our results 
support the dominance of gluonic effects, they also suggest that the phase boundary
is likely to be first order.

\bigskip
\noindent 
The authors would like to thank Christina Manuel, Rob Pisarski, Krishna Rajagopal, 
Ramamurti Shankar and Thomas Schaefer for useful discussions and comments. 
NE and MS would like to thank the Aspen Center for Physics for their
hospitality while this work was completed.
This work was supported in part under DOE contracts 
DE-FG02-91ER40676 and DE-FG06-85ER40224.



\vskip 1 in
\baselineskip=1.6pt
\begin{thebibliography}{99}
%
%
\def\np#1#2#3{  {Nucl. Phys. #1} (19#3) #2}
\def\pl#1#2#3{  {Phys. Lett. #1} (19#3) #2}
\def\pr#1#2#3{   {Phys. Rev. #1} (19#3) #2}
\def\prep#1#2#3{ {Phys. Rep. #1} (19#3) #2)}
\def\prl#1#2#3{ {Phys. Rev. Lett. #1} (19#3) #2}
%

\bibitem{FM} Freedman and McLerran, \pr{D16}{1130}{77};
\pr{D16}{1147}{77};\pr{D16}{1169}{77}.

\bibitem{BCSrev} See, for example, N. Ashcroft and N.D. Mermin,
Solid State Physics, Saunders College Publishing (1976).

\bibitem{FLRG} G. Benfatto and G. Gallavotti, J. Stat. Phys. 59, 541 (1990);
\pr{C42}{9967}{90}; R. Shankar, Physica A177, 530 (1991); 
Rev. Mod Phys. 66, 129 (1993); J. Polchinski, in Proceedings of the 1992 TASI,
eds. J. Harvey and J. Polchinski (World Scientific, Singapore 1993).

\bibitem{BL} D. Bailin and A. Love, \np{B190}{175}{81};\np{B190}{751}{81};
\np{B205}{119}{82};\prep{107}{325}{84}. 

\bibitem{instanton}  R. Rapp, T. Schafer, E.V. Shuryak and M. Velkovsky,
 \prl{81}{53}{98}, hep-ph/9711396;
 M. Alford, K. Rajagopal and F. Wilczek,
\pl{B422}{247}{98}, hep-ph/9711395; J. Berges and K. Rajagopal,  
hep-ph/9804233.

\bibitem{CFL}
M. Alford, K. Rajagopal and F. Wilczek, hep-ph/9804403. 



\end{thebibliography}
\end{document}